\documentclass[%
 aip,
 jcp,%
 amsmath,amssymb,
 preprint,%
]{revtex4-1}

\usepackage{CJK}
\usepackage{amsmath}
\usepackage{graphicx}
\usepackage{dcolumn}
\usepackage{bm}

\draft 

\begin{document}
\begin{CJK*}{UTF8}{}


\title{Application of the R-matrix method to Photoionization of Molecules}


\author{Motomichi TASHIRO (\CJKfamily{min}田代基慶)}
\email[E-mail:]{tashiro@ims.ac.jp}

\affiliation{Institute for Molecular Science, 
Nishigo-Naka 38, Myodaiji, Okazaki 444-0867, Japan}


\date{\today}

\begin{abstract}
The R-matrix method has been used for theoretical calculation of electron 
collision with atoms and molecules for long years. The method was also 
formulated to treat photoionization process, however, its application has been 
mostly limited to photoionization of atoms. 
In this work, we implement the R-matrix method to treat molecular photoionization problem 
based on the UK R-matrix codes. 
This method can be used for diatomic as well as polyatomic molecules, 
with multi-configurational description for electronic states of both 
target neutral molecule and product molecular ion. 
Test calculations were performed for valence electron photoionization of 
nitrogen (N$_2$) as well as nitric oxide (NO) molecules. 
Calculated photoionization cross sections and asymmetry parameters agree 
reasonably well with the available experimental results, suggesting 
usefulness of the method for molecular photoionization. 

\end{abstract}



\maketitle
\end{CJK*}


\section{Introduction}

Photoionization of molecule has been studied for long year because of its 
importance in modeling and understanding upper atmosphere, interstellar clouds 
and industrial plasma. 
Photoionization process has also been used to understand nature of molecular 
electronic states as well as dynamics of nuclei on 
excited electronic state \cite{ISI:000237668700019}.  
In addition to the conventional measurement of integral cross section and 
photoelectron asymmetry parameter for randomly oriented molecule, 
recent developments of experimental technique have made it possible to 
extract photoelectron angular distribution from oriented molecule \cite{ISI:000185093900015},  
perform accurate measurement of inner-shell photoionization cross section \cite{ISI:000184012300006}.  

Several theoretical methods have been developed for treating 
photoionization of molecule, based on 
the Schwinger variational principle\cite{ISI:A1986AXU6700001}, 
random phase approximation\cite{ISI:000075052000018,ISI:000185059500011}, 
time-dependent density functional theory\cite{ISI:000246176100024}, 
complex basis function\cite{ISI:A1985ABC5000012,ISI:000084887600014}, 
Stieltjes imaging method\cite{ISI:A1973R260400016}, and so on. 
These methods have successfully reproduced or predicted experimental results for 
photoionization of small to medium sized molecules in the gas phase. 

The R-matrix method, originally developed for nuclear reaction\cite{PhysRev.72.29}, 
has been applied for accurate calculation of cross section for electron collision 
with atoms, molecules and ions with great success\cite{ISI:A1995TJ53300011,Mo98,Te99,Go05,Bu05}. 
As the other electron scattering theories such as the Schwinger variational 
method\cite{ISI:A1986AXU6700001} and the Kohn variational method\cite{ISI:A1993LY60300023}, 
the R-matrix method was also formulated 
to treat photoionization of atoms and molecules\cite{ISI:A1975AX03900020}. 
In addition to the single-photon ionization process, the method was also modified to 
treat multi-photon ionization process\cite{ISI:A1991FB06900005,ISI:000085201600008}. 
There have been many applications of the R-matrix method to 
photoionization of atoms \cite{ISI:A1975AX03900020,ISI:A1995TJ53300011}, however, 
its application to molecular photoionization has been fairly limited, 
with only a few reports on hydrogen molecule\cite{ISI:A1986C151500001,ISI:000169665700007}. 
The ab initio R-matrix method can reproduce the existing experimental 
cross sections very well, as has been demonstrated in the previous 
electron-molecule scattering calculations\cite{2006PhRvA..74b2706T,2007PhRvA..75a2720T}. 
Thus, the method is expected to work efficiently in 
molecular photoionization problem as well. 
Introduction of a different method, the R-matrix method, may be useful 
when various theoretical methods are compared with experiment. 
In the R-matrix treatment of photoionization, it is possible to represent electronic 
states of target molecular ion by multi-configurational wavefunctions, 
as it has been in the R-matrix calculations on electron molecule collisions
\cite{ISI:A1995TJ53300011,Mo98,Te99,Go05,Bu05,2006PhRvA..74b2706T,2007PhRvA..75a2720T}. 
This multi-configurational representation of wavefunction may be good for description of 
satellite or resonance states with multi-configurational character, 
as well as introducing correlation in the initial and final states. 

In this work, we implement the R-matrix method of molecular photoionization 
based on the work of Burke and Taylor\cite{ISI:A1975AX03900020} originally proposed 
for atomic photoionization. 
Our procedure is roughly divided into two stages. In the first stage, we perform 
R-matrix calculations for electron collision with molecular ion, and obtain 
eigenvalues and eigenvectors of the R-matrix eigenstates by diagonalizing the electronic 
Hamiltonian inside the R-matrix sphere. In the second stage, the initial and final state 
wavefunctions of photoionization are constructed as linear combinations of 
the R-matrix eigenstates obtained in the first stage, then the transition dipole 
moments between these initial and final state are obtained for calculation of 
photoionization cross section and asymmetry parameter. 
For calculation in the first stage, we use the polyatomic version of the UK R-matrix 
codes developed by Morgan et al.\cite{Mo98} The UK R-matrix codes has been 
successfully applied to many electron-molecule scattering problems in the past. 
In this work, some modifications are made to the UK R-matrix codes, e.g., to 
enable transition dipole moment calculation between the R-matrix eigenstates. 
In order to verify reliability and accuracy of the R-matrix method for 
molecular photoionization, cross sections and asymmetry parameters are calculated for valence 
electron photoionization of nitrogen (N$_2$) and nitric oxide (NO) molecules randomly 
oriented in the gas phase, and the results are compared with existing experimental and 
theoretical data. 
N$_2$ has been benchmark molecule for various theories of photoionization, 
and will be also good for the first test calculation of the R-matrix method for molecular 
photoionization. 
In contrast to N$_2$ molecule, number of calculations on photoionization of NO molecule is 
limited\cite{ISI:A1996UP20200016}. 
We selected NO molecule as the second test case, to see 
capability of the method to treat this open-shell molecule 
with $\Pi$ ground state.

\section{Theoretical method}

In this section, we describe the procedure to obtain molecular photoionization 
cross section and asymmetry parameter using the R-matrix method. 
Our discussion mainly follows Chandra\cite{ISI:A1986E175400011, ISI:A1987J389100013} and 
Burke and Taylor\cite{ISI:A1975AX03900020}. 
Many equations in this section already appear in their papers, however, 
they are shown here for reader's convenience. 

Differential cross section for photoionization of molecule randomly 
oriented in space can be written as,\cite{ISI:A1986E175400011, ISI:A1987J389100013} 
\begin{equation}
\frac{d\sigma}{d\hat{\bm k'}} = \frac{3}{4} \left( \frac{e^2}{\alpha E_r} \right)^2 \sum_{L} A_L \left( k \right) 
P_{L} \left( \cos \theta' \right),
\label{eq1}
\end{equation}
where $\hat{\bm k'}$ represents direction of a photoelectron with wavenumber $k$ in laboratory frame, 
$E_r$ is photon energy, 
$P_{L}$ is a Legendre polynomial, 
$\theta'$ is the angle between the direction of photoelectron and the polarization vector 
of the incident photon. Here we assume that the photon beam is linearly polarized. 
The expansion coefficient $A_L(k)$ is represented as
\begin{eqnarray}
A_L \left( k \right) &=& \left( 2L+1 \right) 
\begin{pmatrix} 1 & 1 & L \\ 0 & 0 & 0 \end{pmatrix}
\sum_{l_f m_f \lambda_r} \sum_{l'_f m'_f \lambda'_r} \left( -i \right)^{l_f-l'_f} 
e^{i\left( \sigma_{l_f} - \sigma_{l'_f} \right)} (-1)^{m_f+\lambda_r}
\left( 2l_f+1 \right)\left( 2l'_f+1 \right) \notag \\
&& \times 
\begin{pmatrix} l_f & l'_f & L \\ 0 & 0 & 0 \end{pmatrix}
\begin{pmatrix} l_f & l'_f & L \\ -m_f & m'_f & \epsilon \end{pmatrix}
\begin{pmatrix} 1 & 1 & L \\ \lambda_r & -\lambda'_r & -\epsilon \end{pmatrix}  
M_{l_fm_f}^{-} \left( \lambda_r \right) M_{l'_fm'_f}^{- *} \left( \lambda'_r \right),
\label{eq2}
\end{eqnarray}
where $l_f$, $l'_f$, $m_f$ and $m'_f$ specify angular quantum number of the photoelectron 
associated with the final electronic state $f$ of the product molecular ion, $\epsilon$ 
equals to $m_f-m'_f$, $\sigma_{l_f}$ is the Coulomb phase, and 
$M_{l_fm_f}^{-} \left( \lambda_r \right)$ represents transition dipole matrix element 
between the initial state $\Psi_{i}$ and the final state $\Psi_{l_fm_f}^{f-}$, 
\begin{equation}
M_{l_fm_f}^{-} \left( \lambda_r \right) = A \langle \Psi_{l_fm_f}^{f-} | 
\sum_{s=1}^{n_e} \hat{\bm e}_{\lambda_r} \cdot {\bm r_s} |  \Psi_{i}  \rangle.
\label{eq3}
\end{equation}
Here $\hat{\bm e}_{\lambda_r}$ is the unit vector describing polarization $\lambda_r$ of the incident 
photon in molecular frame, $n_e$ is number of electrons in the system with ${\bf r}_s$ being 
coordinates of the $s$th electron. 
In this work, we employ dipole length approximation. Thus,  
the proportional coefficient $A$ becomes\cite{ISI:A1986E175400011} 
$\left( 4\pi \alpha^3 E_r^3 / 3 e^4 \right)^{1/2}$ 
with $\alpha$ being the fine structure constant. 
As is well known, the differential photoionization cross section in Eq. (\ref{eq1}) can be simplified 
in the form, 
\begin{equation}
\frac{d\sigma}{d\hat{\bm k'}} = \frac{\sigma_{tot.}}{4\pi} 
\left[ 1 + \beta P_2 \left( \cos \theta' \right) \right],
\label{eq4}
\end{equation}
with the integrated cross section,
\begin{equation}
\sigma_{tot.} = \pi \left( \frac{e^2}{\alpha E_r} \right)^2  
\sum_{l_fm_f\lambda_r} \left| M_{l_fm_f}^{f-} \left( \lambda_r \right) \right|^2, 
\label{eq5}
\end{equation}
and the asymmetry parameter,
\begin{equation}
\beta = \frac{A_{L=2} \left( k \right)}{A_{L=0} \left( k \right)}.
\label{eq6}
\end{equation}

Following Burke and Taylor\cite{ISI:A1975AX03900020}, we evaluate the transition dipole 
matrix elements in Eq. (\ref{eq3}) using the R-matrix method \cite{Mo98,Te99,Go05,Bu05} 
for electron-molecule collision.  
In this method, configuration space is divided by two regions 
according to the distance $r_{n_e}$ of the scattering electron, i.e., photoelectron, and  
the center of mass of the molecular ion having $(n_e-1)$ electrons. 
In the inner region, defined by condition $r_{n_e} < a$, 
the $n_e$-electron wave functions of the total system 
are represented by $(n_e-1)$-electron wave functions of molecular ion augmented by 
diffuse functions which describe a scattering electron, 
\begin{equation}
\Phi_k = \textrm{$\mathcal{A}$} \sum_{ij} \Bar{\phi}_{i} \left(x_{1} ,\dots, x_{n_e-1} \sigma_{n_e}\right) 
u_{j} \left({\bf r}_{n_e} \right) a_{ij} 
+ \sum_{q} X_{q} \left(x_{1} \dots x_{n_e} \right) b_{q}, 
\label{eq7}
\end{equation}
where $\mathcal{A}$ is an antisymmetrization operator,  
$x_{i}$ represents spacial and spin coordinates of the $i$th electron, 
$\Bar{\phi}_{i}$ are the eigenstate of the total spin and its z component constructed from 
the $(n_e-1)$-electron wave functions of the molecular ion $\phi_{i}$ and 
the spin function of the scattering electron, 
$u_{j}$ are continuum orbitals representing wave functions of the scattering electron,
and $X_{q}$ are bound $n_e$ electron wave functions composed 
of the target molecular orbitals and the extra target virtual orbitals, 
the coefficients $a_{ij}$ and $b_{q}$ are determined by diagonalization 
of $H_{n_e} + L_{n_e}$ where $H_{n_e}$ is the electronic Hamiltonian and 
$L_{n_e}$ is a Bloch operator \cite{1957NucPh...4..503B, ISI:A1994NV16900022} 
accounting for surface term. 
In the outer region $r_{n_e} > a$, the problem is reduced to single 
electron scattering, ignoring exchange of the scattering electron with 
the electrons of the molecular ion. Interaction of the scattering electron and the
molecular ion is considered through static multipolar interaction potentials which 
introduce inter-channel couplings. 
The R-matrix eigenstates obtained by diagonalization of the Hamiltonian in the inner region 
are converted to the R-matrix $R_{ij}$ at the boundary $r_{n_e} = a$ as, 
\begin{equation}
R_{ij} \left( E \right) = 
\frac{1}{2a} \sum_k \frac{w_{ik} \left( a \right)w_{jk} \left( a \right)}{E_k - E},
\label{eq8}
\end{equation}
where $E_k$ is eigenvalue of $\Phi_k$, $E$ is the energy of the total system, 
and the boundary amplitudes $w_{ik}$ is given by 
\begin{equation}
w_{ik} \left( r \right) = \langle \Bar{\phi}_i Y_{l_im_i} | \Phi_k \rangle. 
\label{eq9}
\end{equation}
For photoionization problem, the final states $\Psi_{l_fm_f}^{f-}$ 
in Eq. (\ref{eq3}) are expanded by the R-matrix eigenstates $\Phi_k$ as\cite{ISI:A1975AX03900020}, 
\begin{equation}
\Psi_f^- = \sum_k A_{kf} \Phi_k.
\label{eq10}
\end{equation}
Here $\Psi_{l_fm_f}^{f-}$ is denoted as $\Psi_{f}^{-}$ for simplicity. 
Using the relation \cite{ISI:A1975AX03900020,ISI:A1971J033100003} which holds for the R-matrix 
amplitudes and the final state wave functions, the expansion coefficients can be written as,
\begin{equation}
A_{kf} = \frac{1}{2a \left( E_k - E_f \right)} 
\sum_j w_{jk} \left( a \right) \left( a \frac{dy_{jf}}{dr} - b y_{jf} \right)_{r=a},
\label{eq11}
\end{equation}
where $E_f$ is the energy of the final state, which is equal to $E_i + E_r$ with 
the initial state energy $E_i$ and the photon energy $E_r$. 
The parameter $b$ is related to the logarithmic derivative of the scattering electron 
wavefunction at the R-matrix boundary and is set to zero in this work. 
The radial function $y_{jf}$ is defined as
\begin{equation}
y_{jf} = \langle \Bar{\phi}_j Y_{l_jm_j} | \Psi_f^- \rangle. 
\label{eq12}
\end{equation}
We have to determine the radial functions and their derivatives in Eq. (\ref{eq11}) 
to evaluate the final state wave functions. 
In the outer region, the radial functions $y_{jf}$ satisfy the differential 
equations \cite{ISI:A1975AX03900020,ISI:A1971J033100003},

\begin{equation}
\left( \frac{d^2}{dr^2} - \frac{l_i \left( l_i + 1 \right) }{r^2} + \frac{2z}{r} + k_j^2 \right)
y_j \left( r \right) = 
2 \sum_{k=1}^{n} V_{jk} \left( r \right) y_k \left( r \right),
\label{eq13}
\end{equation}
where $n$ is the number of channels considered in the R-matrix model,
$z$ is the net charge of the molecular ion, 
$k_j$ is the wavenumber of the electron in the $j$th channel 
and $V_{jk} \left( r \right)$ represents multipole potential.\cite{ISI:A1971J033100003}   
When $n_a$ channels are open, we have $n+n_a$ independent solutions of Eq. (\ref{eq13}) 
with the asymptotic boundary conditions,

\begin{align}
v_{ij} \left( r \right) &\underset{r\to\infty}{\sim} k_i^{-1/2} \sin \theta_i \delta_{ij} 
&&i=1,..,n, ~ ~ ~ j=1,..,n_a        \notag \\
v_{ij} \left( r \right) &\underset{r\to\infty}{\sim} k_i^{-1/2} \cos \theta_i \delta_{ij-n_a} 
&&i=1,..,n, ~ ~ ~ j=n_a+1,..,2n_a   \notag \\
v_{ij} \left( r \right) &\underset{r\to\infty}{\sim} \exp \left( - \left| k_i \right| r \right)
\delta_{ij-n_a}
&&i=1,..,n, ~ ~ ~ j=2n_a+1,..,n+n_a
\label{eq14}
\end{align}
where
\begin{equation}
\theta_i = k_i r - \frac{1}{2} l_i \pi - \eta_i \ln\left( 2 k_i r \right) + 
\arg \Gamma \left( l_i + 1 + i \eta_i \right)
~ ~ ~ i=1,..,n_a
\label{eq15}
\end{equation}
with $\eta_i = -z / k_i$.
The radial functions $y_{jf}$ are expanded by these $n+n_a$ independent solutions $v_{ij}$ as, 
\begin{equation}
y_{jf} = \sum_{k=1}^{n+n_a} v_{jk} x_k.
\label{eq16}
\end{equation}
By inserting Eq. (\ref{eq16}) into the R-matrix relation \cite{ISI:A1971J033100003,ISI:A1975AX03900020}, 
\begin{equation}
y_{jf} \left( a \right) = \sum_{k=1}^{n} R_{jk} \left( a \frac{dy_{kf}}{dr} - b y_{kf} \right)
~ ~ ~ j=1,..,n,
\label{eq17}
\end{equation}
and using the ingoing wave asymptotic conditions of $y_{f}$,
\begin{equation}
\left[ y_{f}^- \right]_{ij} \underset{r\to\infty}{\sim} \sum_{k=1}^{n_a} k_i^{-1/2} 
\left( \sin \theta_i \delta_{ik} + \cos \theta_i K_{ik} \right)
\left[\left( {\bm 1} + i {\bm K} \right)^{-1} \right]_{kj}
~ ~ ~ i,j=1,..,n_a,
\label{eq18}
\end{equation}
with ${\bm K}$ and $K_{ik}$ being K-matrix and its elements, 
we obtain $(n+n_a)$ linear equations for $(n+n_a)$ unknown coefficients $x_k$. 
The expansion coefficients $x_k$ can be determined by solving these equations, and as a result, 
we can evaluate the final state wavefunctions $\Phi_f^-$ using Eqs. (\ref{eq10}) and (\ref{eq11}). 

In this work, the initial state wave function $\Psi_i$ is also expanded by 
the R-matrix eigenstates as described in Burke and Taylor \cite{ISI:A1975AX03900020}, 
\begin{equation}
\Psi_i = \sum_k A_{ki} \Phi_k,
\label{eq19}
\end{equation}
where the coefficient $A_{ki}$ is given by Eq. (\ref{eq11}) with $E_f$ and $y_{jf}$ replaced 
by the energy $E_i$ and the radial function $y_{ji}$ of the initial state. 
The radial functions $y_{ji}$ are expanded by the independent solutions of the 
differential equations (\ref{eq13}) as,
\begin{equation}
y_{ji} = \sum_{k=1}^{n} v_{jk} x_k.
\label{eq20}
\end{equation}
Since we are treating bound initial state, all channels are closed, $n_a=0$, and 
all independent solutions decay to zero as $r$ approaches infinity. 
By substituting Eq. (\ref{eq20}) into the R-matrix relation of Eq. (\ref{eq17}), 
$n$ equations for $n$ unknown coefficients $x_k$ are obtained. 
These equations are only solved at discrete energies of $E_i$, where the lowest of them 
corresponds to the ground state energy of the neutral molecule. 

\section{Application of the method to photoionization of N$_2$ and NO molecules}

\subsection{Detail of the calculations}

Cross sections and asymmetry parameters for valence electron photoionization of 
N$_2$ and NO molecules were calculated based on the method described in the previous 
section. The R-matrix eigenstates and amplitudes were obtained by 
electron - N$_2^+$ and NO$^+$ scattering calculations using a modified version of the 
polyatomic programs in the UK molecular R-matrix codes \cite{Mo98}. 
We used the fixed nuclei approximation with internuclear distances 
2.068 a$_0$ for N$_2$ and 2.175 a$_0$ for NO, which are the equilibrium values 
of the ground electronic states of N$_2$ and NO. 

For the electron - N$_2^+$ R-matrix scattering calculation, we employed two R-matrix models; 
a single-channel SCF target model and a multi-channel CI target model including 80 N$_2^+$ 
electronic states. The cc-pVTZ atomic basis set \cite{1989JChPh..90.1007D} was used to 
describe molecular orbitals in both models. 
In the CI target model, full valence complete active space was employed 
to generate configuration state functions for the N$_2^+$ electronic states. 
The molecular orbitals (MOs) in this active space were obtained by the state-averaged 
complete active space self consistent field (CASSCF)\cite{ISI:A1985AGJ3300004,ISI:A1985AJG5000041} 
calculation using molpro program package\cite{MOLPRO}. 
The state-averaging was performed over the lower 28 electronic states of N$_2^+$ ion, 
then the CASCI wave functions of the remaining 52 electronic states were 
constructed from this CASSCF molecular orbital set. The CASSCF ionization potentials of 
the lower 9 N$_2^+$ electronic states are listed in Table \ref{tab1}, 
where the N$_2$ ground state energy was obtained by the electron - N$_2^+$ scattering 
calculation, as described at the end of the last section. 
In addition to these CASSCF MOs for the N$_2^+$ valence electronic states, 
we included 3 extra virtual orbitals for each irreducible representation of 
the D$_{2h}$ symmetry. 
In order to represent the scattering electron, we included diffuse
Gaussian functions up to $l$ = 5, with 12 functions for $l$ = 0 and 1, 8 functions 
for $l$ = 2 and 3, and 5 functions for $l$ = 4 and 5.  
Exponents of these diffuse Gaussians were obtained by the GTOBAS 
program \cite{Fa02} in the UK R-matrix codes.  
To prepare an orthogonal MO set used in Eq. (\ref{eq7}), 
these Gaussian functions were orthogonalized against the valence and extra 
virtual MOs obtained by the CASSCF calculation. 
The procedure for construction of the 14-electron configurations in Eq. (\ref{eq7}) 
is almost the same as we did in the previous work of electron - N$_2$ 
scattering calculation \cite{2007PhRvA..75a2720T}, except the number of electron in the system,  
and is not repeated here. 
Radius of the R-matrix sphere was chosen to be 10 a$_0$ in our calculations.
The R-matrix calculations were performed for the singlet $A_g$, $B_{1u}$, $B_{2u}$ and $B_{3u}$ 
symmetries of the e+N$_2^+$ system, where the $A_g$ result was used for the ground state of the 
neutral molecule whereas the other symmetries were used to construct the final state 
wavefunctions. 

Detail of the electron - NO$^+$ R-matrix scattering calculation is similar to 
the electron - N$_2^+$ scatterings. In this case, we performed 
a single-channel SCF target calculation and a multi-channel CI target 
calculation with 60 NO$^+$ electronic states. 
The state-averaged CASSCF calculation with cc-pVTZ basis set was performed 
for the lowest 16 electronic states of NO$^+$, then the CASCI wave functions 
were constructed for the other states. The ionization potentials of the lowest 8 
NO$^+$ electronic states are shown in Table \ref{tab2}. 
Radius of the R-matrix sphere and diffuse Gaussian functions 
for the scattering electron are the same as in the e-N$_2^+$ calculation.
In this e - NO$^+$ calculation, we included 4 extra virtual orbitals for 
each irreducible representation of the C$_{2v}$ symmetry. 
The R-matrix calculations were performed for the doublet $A_1$, $A_{2}$ and $B_{1}$ 
symmetries of the e+NO$^+$ system, where the $B_1$ result was used to describe 
both the ground state of the neutral molecule and the final state wavefunctions, 
whereas the other symmetries were only used for the final states. 

The wave functions of the initial N$_2 {X}^1 \Sigma_g^+$ and NO$ {X}^2 \Pi$ states 
were expanded by the R-matrix eigenstates taken from the e - N$_2^+$ and e - NO$^+$ 
R-matrix scattering calculations as described at the end of the last section, 
and their expansion coefficients were obtained by the BOUND module of the UK R-matrix codes. 
It was found that the wave functions of these N$_2$ and NO states are dominated by 
the lowest eigenvalue R-matrix eigenstates, i.e., expansion coefficients of 
the lowest eigenvalue states are almost unity whereas they are less than $10^{-4}$ for 
the other R-matrix eigenstates. 
Thus, we just substituted the initial state wave functions in Eq. (\ref{eq3}) 
by these lowest eigenvalue R-matrix eigenstates. 
Note that the above procedure means that the ground state wavefunction of the 
neutral molecule is described by the MOs of the molecular ion. 
This is approximate method for description of the ground 
state wavefunctions, however, it greatly simplifies evaluation of 
the transition dipole moments, since the same set of MOs 
is used in the initial and the final state wavefunctions. 

In order to obtain the transition dipole matrix elements of Eq. (\ref{eq3}), 
we evaluated the dipole matrix elements between the R-matrix eigenstates. 
Then the expansion coefficients for the final state wave functions 
were calculated using Eqs. (\ref{eq11}) and (\ref{eq16}). 
In principle, accurate solutions of differential equations (\ref{eq13}) with 
proper boundary conditions (\ref{eq14}) have to be used for Eq. (\ref{eq16}). 
Such solutions are usually prepared by the asymptotic expansion 
\cite{PhysRev.126.147, ISI:A1976BP63700027} and 
inward integration to the matching point $r = a$.  
However, the asymptotic expansion is not so accurate for small $r$ and 
near the ionization thresholds, where we have 80 such thresholds for N$_2$ 
photoionization and 60 for NO molecule in our R-matrix models. 
Also, inward integration is unstable due to exponential growing 
of closed channel components. 
In this work, we approximated $v_{ij}$ by the Coulomb functions 
ignoring multipole potentials of the molecular ions. 
Based on the calculated transition dipole moments, photoionization cross sections 
and asymmetry parameters were evaluated using Eqs. (\ref{eq2}), (\ref{eq5}) 
and (\ref{eq6}).

\subsection{Results and Discussion}

\subsubsection{Ionization Potentials}

In Table \ref{tab1}, the CASSCF ionization energies of N$_2$ molecule are shown 
for the lower N$_2^+$ ion states with the experimental values\cite{ISI:A1992JX83400040}. 
Our results agree reasonably well with the experimental ionization energies, 
though ours are 0.5 - 1.0 eV higher. 
The ionization potentials of NO molecule are also shown for the lowest 8 NO$^+$ ion states 
in Table \ref{tab2}. 
Since the equilibrium bond distances of some excited states are longer than that of 
the neutral NO ground state \cite{ISI:A1979HQ74600016}, energetic order of 
the ${A}^1 \Pi$, ${A'}^1 \Sigma^-$ and ${W}^1 \Delta$ states in our results is different  
from that in the adiabatic experimental results. 
Nevertheless, our IPs compare reasonably well with the experiment, with 
deviations less than 1 eV. 
Note that the IPs in Tables \ref{tab1} and \ref{tab2} were extracted from 
the ionic and neutral energies used in the R-matrix photoionization 
calculations. Since we did not attempt to adjust these energies to the 
experimental IPs, the ionization thresholds in our cross sections and 
asymmetry parameters are shifted relative to the correct experimental values. 

\subsubsection{N$_2$}

\begin{figure}
 \includegraphics[scale=1.5]{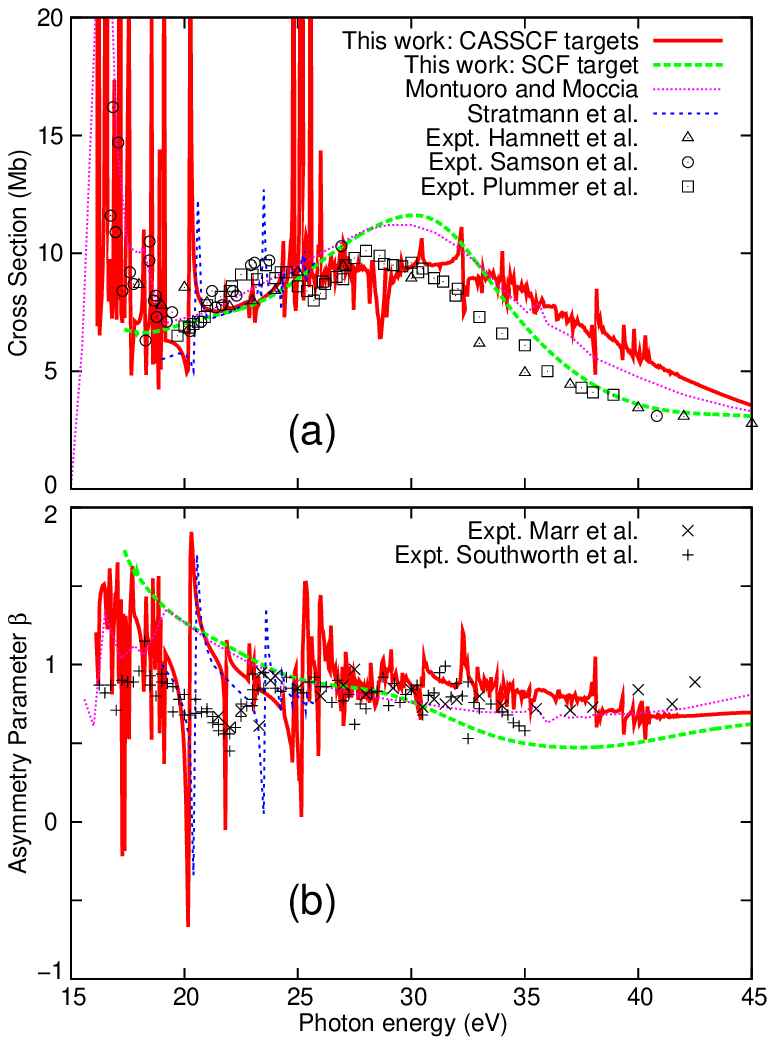}%
 \caption{\label{fig1} 
 Photoionization cross section (a) and asymmetry parameter (b) for ionization of the 
 N$_2^+ {X}^2 \Sigma_g^+$ state. The cross sections are shown in unit of mega barn (Mb), 
 equal to $10^{-22} {\rm m}^2$. 
 Our CASSCF target results and SCF target results are represented as thick full lines and 
 thick dashed lines, respectively. The previous theoretical results of Montuoro and 
 Moccia \cite{ISI:000185059500011} and Stratmann et al.\cite{ISI:A1995RA43100054} are shown 
 as thin dotted and thin dashed lines, respectively. 
 Experimental data included in the panel (a) are taken from Hamnett et al.\cite{ISI:A1976BR99500003}, 
 Samson et al.\cite{ISI:A1977DN87500024} and Plummer et al\cite{ISI:A1977DK70900027}.
 Experimental data in the panel (b) are taken from Marr et al.\cite{ISI:A1979GG10800013}
 and Southworth et al.\cite{ISI:A1986AYM1800027} 
   }
\end{figure}

In Fig. \ref{fig1}, cross section and asymmetry parameter for photoionization of 
the N$_2 {X}^1 \Sigma_g^+$ state leading to the N$_2^+ {X}^2 \Sigma_g^+$ state are shown. 
In the figure, our CASSCF results and SCF results are compared 
with the previous theoretical results of Montuoro and Moccia \cite{ISI:000185059500011} 
obtained by the K-matrix method with interacting channels random phase approximation  
and Stratmann et al.\cite{ISI:A1995RA43100054} obtained by the multi-channel Schwinger 
variational method. 
Also, the experimental cross sections of Hamnett et al.\cite{ISI:A1976BR99500003}, 
Samson et al.\cite{ISI:A1977DN87500024}, Plummer et al\cite{ISI:A1977DK70900027}., 
and the asymmetry parameters of Marr et al.\cite{ISI:A1979GG10800013} and 
Southworth et al.\cite{ISI:A1986AYM1800027} are included in the figure. 
Our SCF cross section is smooth and has a large broad peak around 
30 eV, originated from the $\sigma^*$ shape resonance. 
In contrast, our CASSCF multi-channel cross section has numerous sharp peaks 
originated from two- or many-electron excited resonances as well as 
Rydberg resonances associated with the excited electronic states of N$_2^+$ ion.  
Overall shape of the cross section is roughly similar to the SCF result. 
The CASSCF cross section rises at low energy region around 15-20 eV, which is 
consistent with the experimental result of Hamnett et al. and Samson et al. 
and the previous theoretical cross section of Montuoro and Moccia. 
Below 30 eV, our CASSCF cross section agrees well, 
except for the presence of the sharp peaks, with the experimental results.  
However, it overestimates the experimental values above 30 eV. 
The SCF cross section agrees rather well with the experimental results 
in this high energy region. 
Similar to the calculated cross sections, the CASSCF asymmetry parameter $\beta$ 
has numerous sharp peaks while the SCF asymmetry parameter 
is very smooth. Compared to the SCF result, the CASSCF result agrees better 
with the experimental asymmetry parameters in all energy range in the figure.  
At low energy region below 24 eV, the CASSCF R-matrix calculation overestimates 
the experimental asymmetry parameters, partly due to existence of resonances. 
The cross section and asymmetry parameter of Stratmann et al. 
are available over limited energy range of 19 - 26 eV. In this energy region, 
our CASSCF results are very similar to their results. 

\begin{figure}
 \includegraphics[scale=1.5]{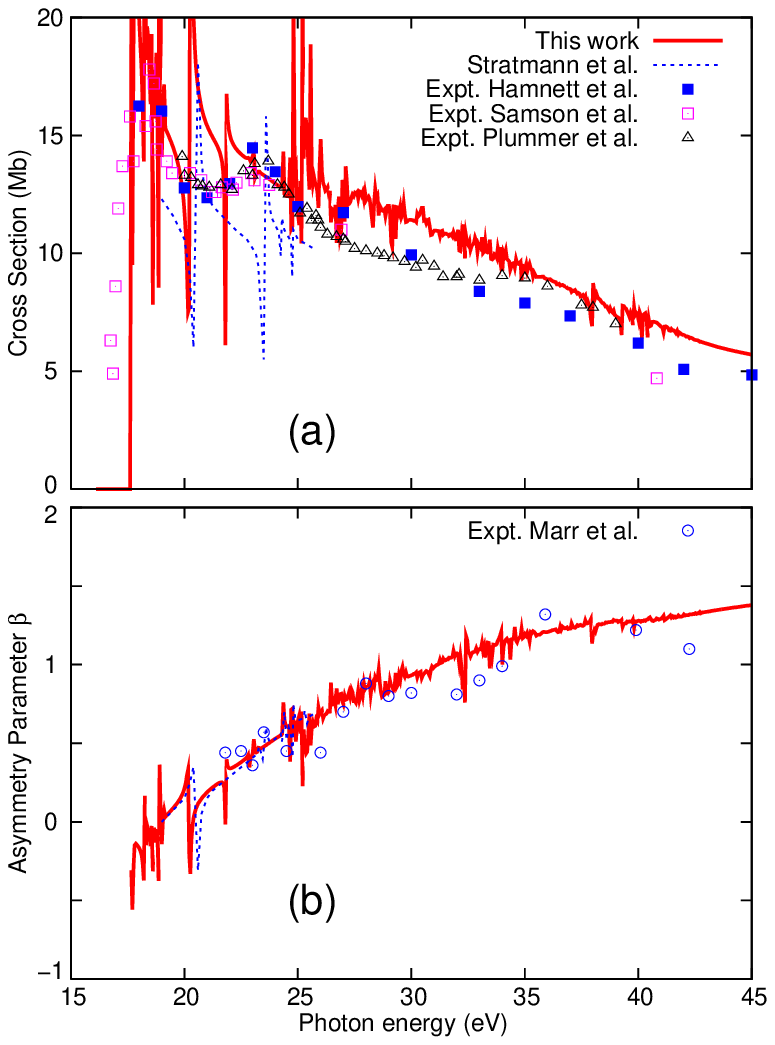}%
 \caption{\label{fig2} 
 Photoionization cross section (a) and asymmetry parameter (b) for ionization of the 
 N$_2^+ {A}^2 \Pi$ state. The other details are the same as in Fig. \ref{fig1}.
 }

\end{figure}

Figure \ref{fig2} shows photoionization cross section and asymmetry 
parameter for the N$_2^+ {A}^2 \Pi$ state, where 
our CASSCF results are compared with the previous theoretical results 
of Stratmann et al.\cite{ISI:A1995RA43100054} and the experimental cross sections of 
Hamnett et al.\cite{ISI:A1976BR99500003}, Samson et al.\cite{ISI:A1977DN87500024}, 
Plummer et al.\cite{ISI:A1977DK70900027}, and the experimental asymmetry parameter 
of Marr et al.\cite{ISI:A1979GG10800013}  Shape of our CASSCF cross section is 
roughly similar to the experimental results. However, magnitude of our 
cross section is slightly larger than the experimental cross sections above 25 eV. 
In case of the asymmetry parameter, our result 
agrees very well, except for the presence of the small peaks, 
with the experimental result of Marr et al. 
The previous multi-channel Schwinger results of Stratmann et al. are 
available between 19 and 26 eV. 
In this energy region, the shape of our cross section is very similar 
to their result, although magnitude of cross section is larger than theirs. 
For the asymmetry parameter, our result almost coincides with the result of 
Stratmann et al. 

\begin{figure}
 \includegraphics[scale=1.5]{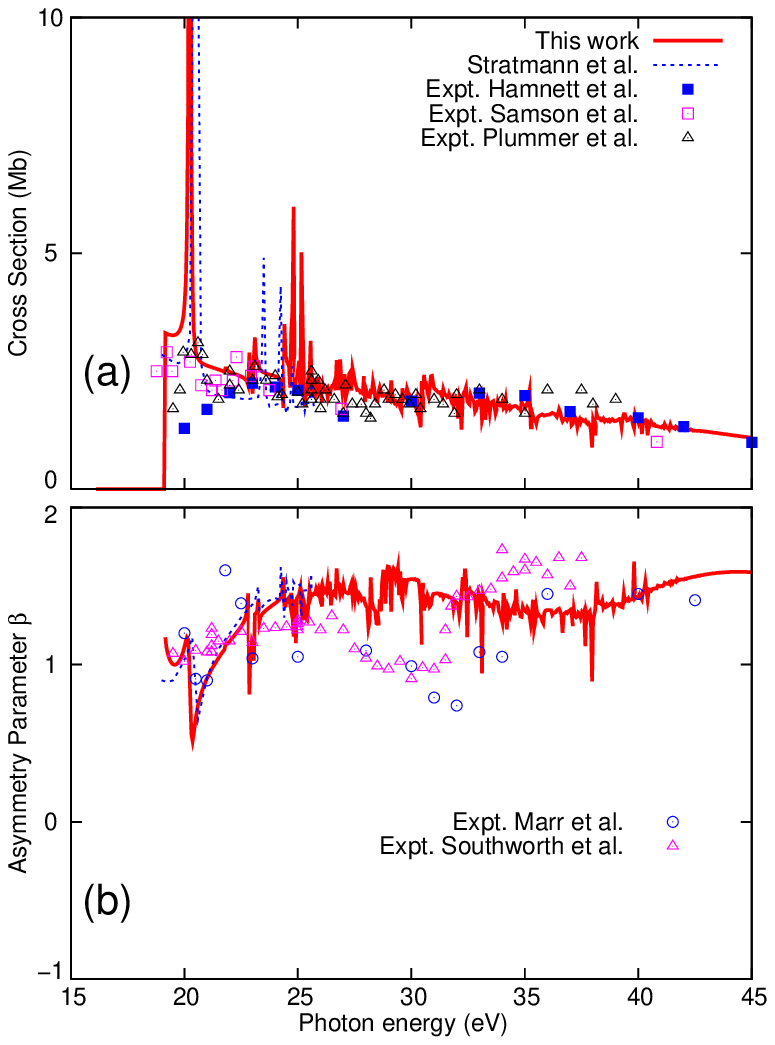}%
 \caption{\label{fig3} 
 Photoionization cross section (a) and asymmetry parameter (b) for ionization of the 
 N$_2^+ {B}^2 \Sigma_u^+$ state. The other details are the same as in Fig. \ref{fig1}.
 }

\end{figure}

In Fig. \ref{fig3}, cross section and asymmetry parameter for 
photoionization leading to the N$_2^+ {B}^2 \Sigma_u^+$ state are shown with 
the available experimental results and the previous theoretical results of 
Stratmann et al.\cite{ISI:A1995RA43100054} As in the photoionization to 
the N$_2^+ {A}^2 \Pi$ state, our cross section is slightly larger than 
the result of Stratmann et al., however, the asymmetry parameter is 
very similar to each other. 
Compared to the experimental results, our cross section looks larger 
near the threshold, yet, agreement is better at higher energies. 
On average, our asymmetry parameter is roughly similar to the experimental 
results, however, the experimental asymmetry parameter has a large dip around 30 eV 
which does not exist in our results. 
In the previous studies \cite{ISI:A1986AYM1800027,ISI:000185059500011}, 
a coupling between the ${X}^2 \Sigma_g^+$ and 
${B}^2 \Sigma_u^+$ channels has been suggested to cause this large dip. 
Although such a coupling is included in our model, 
our asymmetry parameter is rather flat between 20 and 45 eV, 
with slight decrease of magnitude around 35 eV. 

\begin{figure}
 \includegraphics[scale=1.5]{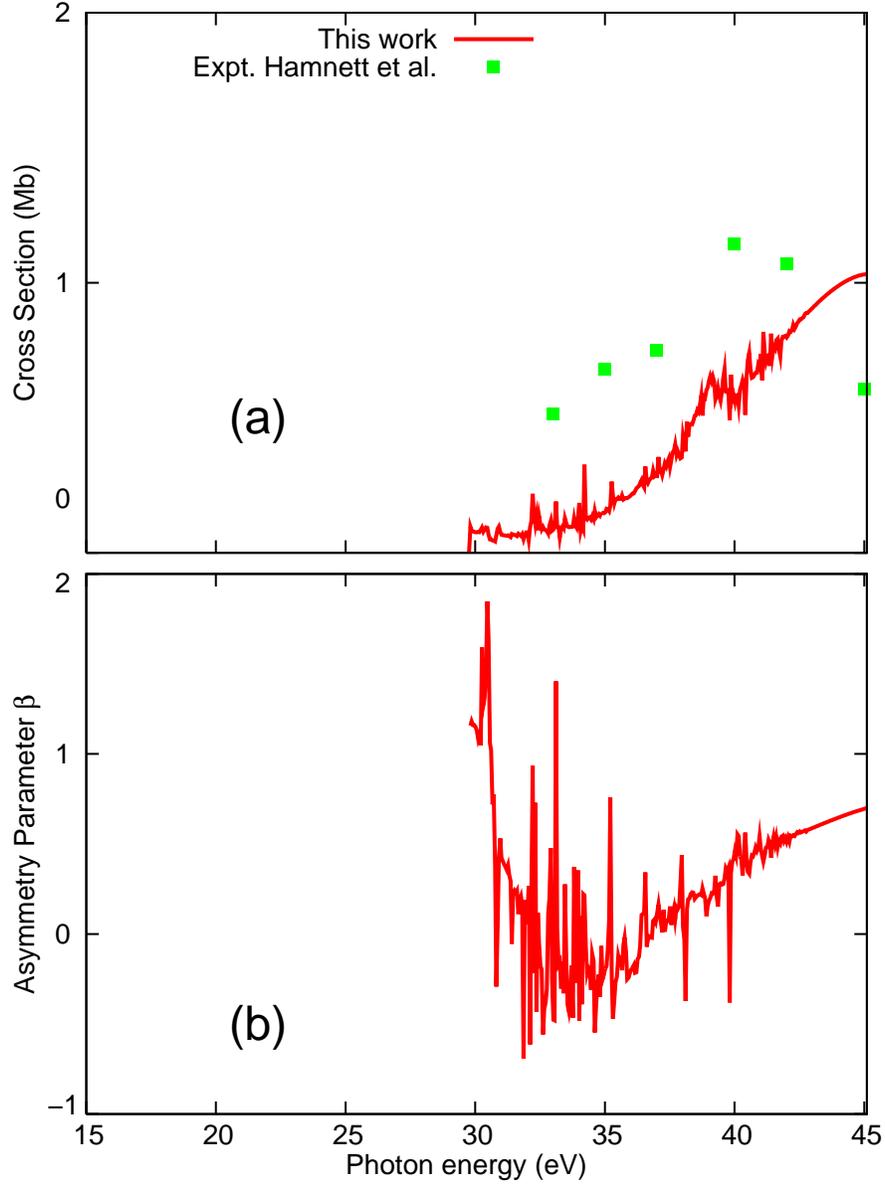}%
 \caption{\label{fig4} 
 Photoionization cross section (a) and asymmetry parameter (b) for ionization of the 
 N$_2^+ {2}^2 \Sigma_g^+$ state. The symbol represents the cross section for 
 the ``Z'' state of N$_2^+$ in Hamnett et al.\cite{ISI:A1976BR99500003}
 The other details are the same as in Fig. \ref{fig1}.
 }
\end{figure}

In general, partial photoionization cross sections for the higher N$_2^+$ ion states are 
very small compared to those for the N$_2^+ {X}^2 \Sigma_g^+$, ${A}^2 \Pi$ and 
${B}^2 \Sigma_u^+$ states. However, the photoionization cross section leading to 
the ${2}^2 \Sigma_g^+$ state is relatively large. The calculated cross section and 
asymmetry parameter are shown in Fig. \ref{fig4}. The ionization energy of 
this N$_2^+ {2}^2 \Sigma_g^+$ state is about 29.65 eV, 
which is close to the threshold of the ``Z'' state mentioned in Hamnett 
et al.\cite{ISI:A1976BR99500003}
Their cross section for the ``Z'' state is also shown in Fig. \ref{fig4}. 
Although agreement is not so good, the magnitudes of the cross sections are roughly 
similar to each other.   
The calculated asymmetry parameter drops from 1.5 at the threshold to -0.5 near 35 eV, 
then it increases to 0.5 at 45 eV. This behaviour differs from the asymmetry parameters 
for the N$_2^+ {X}$, $A$ and $B$ states, probably reflecting the difference in 
the main electronic configurations of the ionic states. 

\subsubsection{NO}

\begin{figure}
 \includegraphics[scale=1.5]{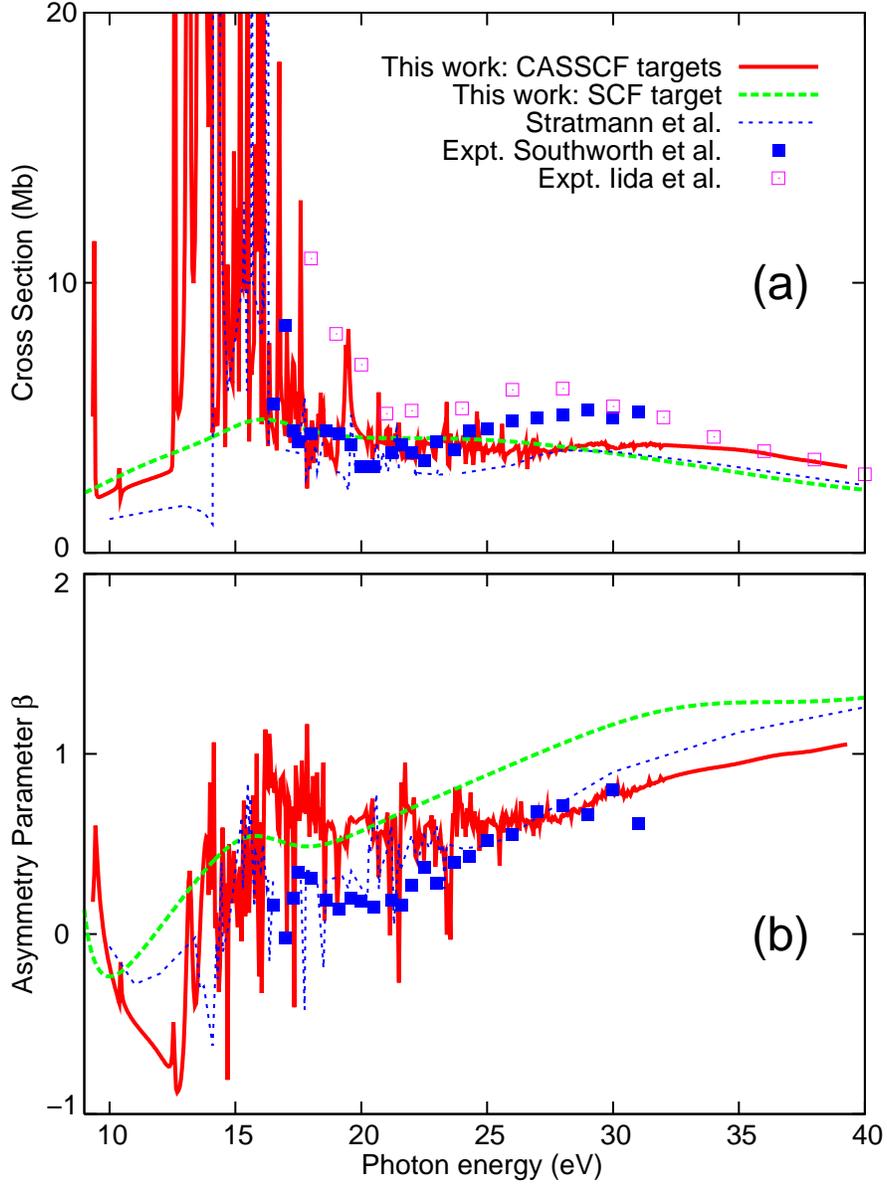}%
 \caption{\label{fig5} 
 Photoionization cross section (a) and asymmetry parameter (b) for ionization of the 
 NO$^+ {X}^1 \Sigma^+$ state. 
 Our CASSCF target results and SCF target results are represented as 
 thick full lines and thick dashed lines, respectively.  
 The previous theoretical results of Stratmann et al.\cite{ISI:A1996UP20200016} are shown as  
 thin dashed lines. Experimental data in the figure are taken from 
 Southworth et al.\cite{ISI:A1982MW46500023} and Iida et al.\cite{ISI:A1986C704300021}
   }
\end{figure}

In Fig. \ref{fig5}, cross section and asymmetry parameter 
for photoionization of the NO$ {X}^2 \Pi$ leading to the NO$^+ {X}^1 \Sigma^+$ state 
are shown with the experimental results of Southworth et al.\cite{ISI:A1982MW46500023} 
and Iida et al.\cite{ISI:A1986C704300021}, and the previous theoretical results of 
Stratmann et al.\cite{ISI:A1996UP20200016} 
Our CASSCF cross section has numerous sharp peaks between 13 and 18 eV, mostly originated 
from Rydberg resonances associated with the excited electronic states of NO$^+$. 
In the other energy region, the magnitude of the cross section is nearly constant value 
of about 5 Mb. 
Compared to the CASSCF result, the SCF cross section is very smooth and flat. 
Although a slight increase of the SCF cross section exists around 16 eV, the height of 
the peak is much smaller than the CASSCF cross section around this energy region. 
The shape of our CASSCF cross section is very similar to the result of 
Stratmann et al, especially above 14 eV. Between 12.5 and 14eV, our cross section has 
several sharp peaks as in the higher energy region. In contrast, there is no such peak in 
the cross section of Stratmann et al. below 14 eV. 
Below 25 eV, our CASSCF cross section agrees well, except for the presence of the sharp peaks, 
with the experimental cross section of Southworth et al. 
Our result, however, slightly underestimates experimental cross section above 25 eV. 
Compared to the result of Iida et al, the magnitude of our CASSCF cross section is generally smaller 
except in the high energy region around 35-40 eV. 
Enhancement of cross section is observed below 20 eV in the experimental results of Southworth 
et al. and Iida et al., which is roughly reproduced by our CASSCF R-matrix calculation.  
However, this feature is not well captured in the SCF R-matrix model.  
The shape of our CASSCF asymmetry parameter is roughly similar to the result of Stratmann et al. 
The asymmetry parameter of Stratmann et al. is rather steeper above 18eV, 
and agrees better with the experimental result of Southworth et al. 
Our SCF asymmetry parameter has a much smoother profile compared to the CASSCF result.  
Its tilt is close to the results of Stratmann et al. and Southworth et al., 
though the magnitude is generally larger. 

\begin{figure}
 \includegraphics[scale=1.5]{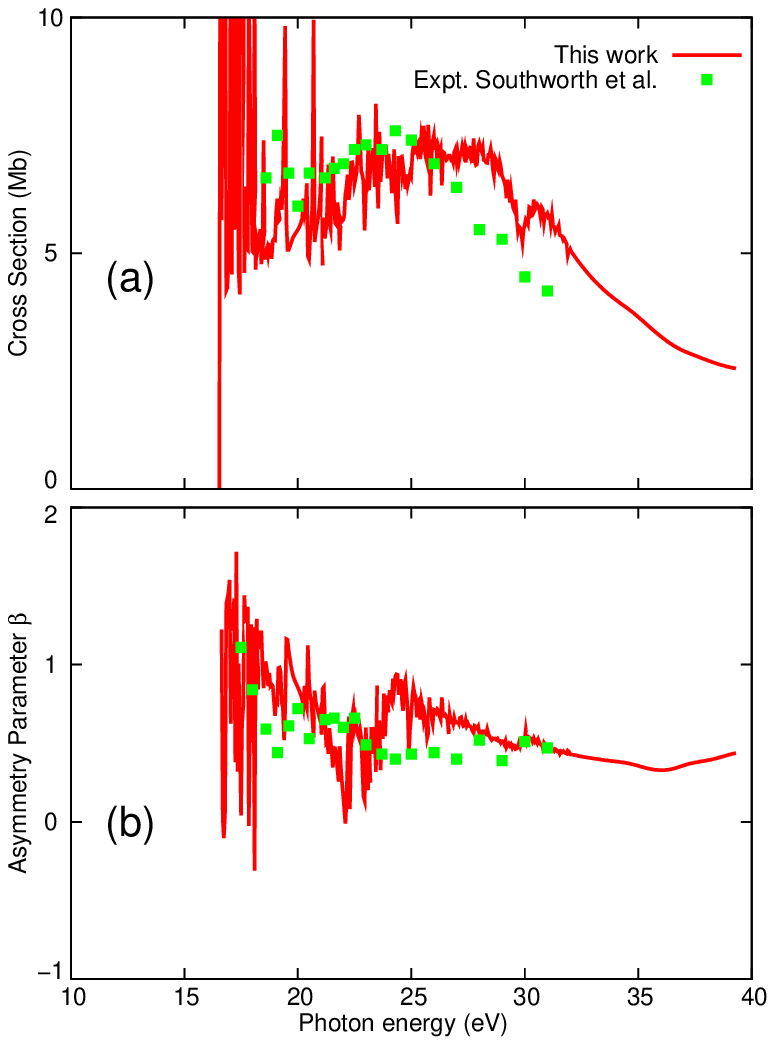}%
 \caption{\label{fig6} 
 Photoionization cross section (a) and asymmetry parameter (b) for ionization of the 
 NO$^+ {b}^3 \Pi$ state. The other details are the same as in Fig. \ref{fig5}.
 }
\end{figure}

Figure \ref{fig6} shows cross section and asymmetry parameter for 
ionization to the NO$^+ {b}^3 \Pi$ state. The experimental results of Southworth 
et al.\cite{ISI:A1982MW46500023} are also included in the figure. 
Near the ionization threshold, our cross section has many peaks as in the cross 
section of the NO$^+ {X}^1 \Sigma^+$ state. The magnitude of the cross section increases 
from 5 Mb to 7 Mb as photon energy increases from the threshold to 27 eV. 
Then it decreases to about 3 Mb at 40 eV. 
Although overall shape of the cross section is similar to the experimental result, 
the broad peak in our result is located 3-4 eV higher in energy compared to the peak 
in the experimental cross section. 
Our asymmetry parameter decreases from 1.5 to 0.4 as energy increases from the 
threshold to 40 eV. This behaviour is roughly similar to the experimental 
result, however, our asymmetry parameter has a dip around 20-25 eV whereas there is a small 
bump in this energy region in the experimental result. 

\begin{figure}
 \includegraphics[scale=1.5]{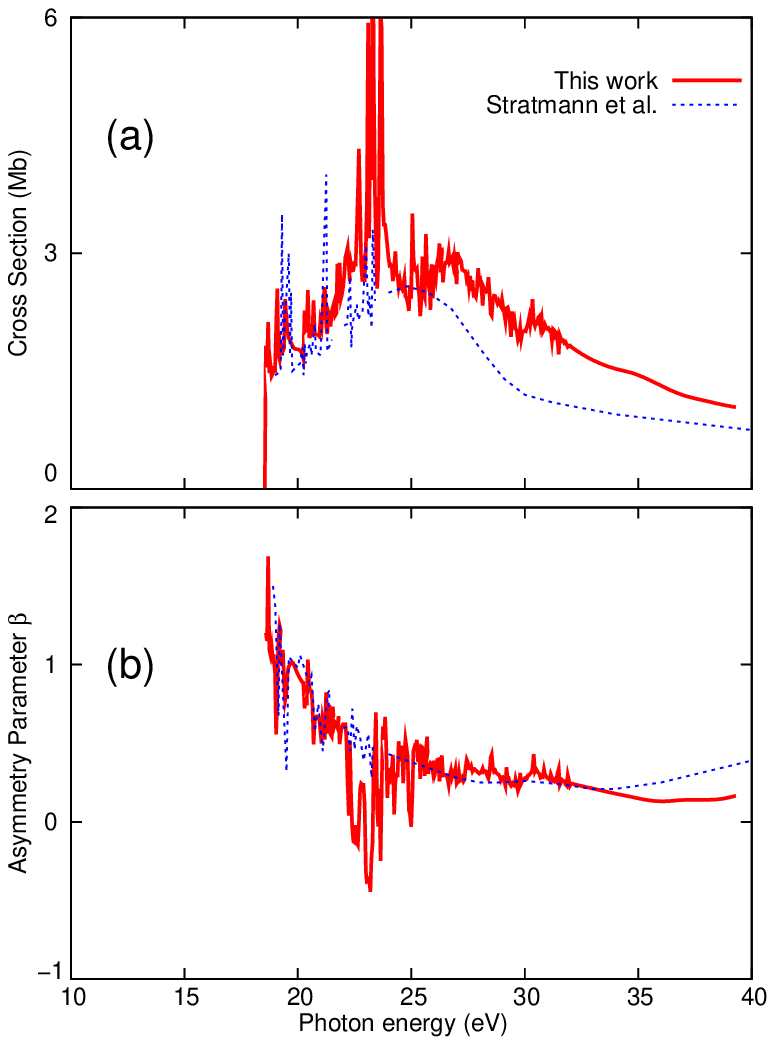}%
 \caption{\label{fig7} 
 Photoionization cross section (a) and asymmetry parameter (b) for ionization of the 
 NO$^+ {A}^1 \Pi$ state. The other details are the same as in Fig. \ref{fig5}.
   }
\end{figure}

In Fig. \ref{fig7}, cross section and asymmetry parameter for photoionization 
to the NO$^+ {A}^1 \Pi$ state are shown with the previous theoretical results of 
Stratmann et al.\cite{ISI:A1996UP20200016} 
Our cross section is generally larger than their result above 22 eV.  
The shapes of the cross sections resemble each other, having a broad peak around 25 eV. 
The asymmetry parameter gradually decreases from 1.5 at the threshold to 0.15 at 40 eV. 
Our result has a dip around 23 eV which does not exist in the asymmetry parameter of 
Stratmann et al., however, the other details are similar. 

\begin{figure}
 \includegraphics[scale=1.5]{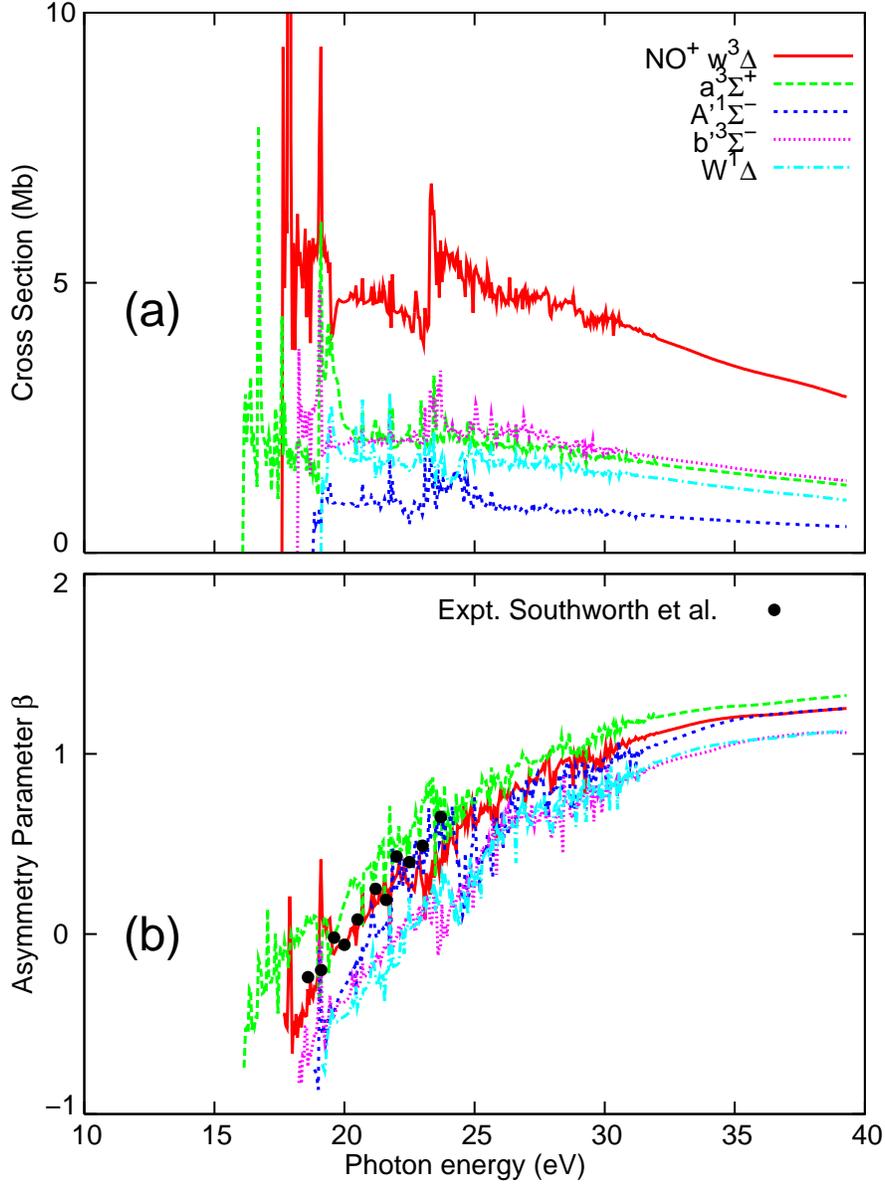}%
 \caption{\label{fig8} 
 Photoionization cross section (a) and asymmetry parameter (b) for ionization of the 
 NO$^+ {w}^3 \Delta$, ${a}^3 \Sigma^+$, ${A'}^1 \Sigma^-$, ${b'}^3 \Sigma^-$ and 
 ${W}^1 \Delta$ states, obtained by the R-matrix calculation in this work. 
 The black dots indicate the experimental asymmetry parameter 
 of Southworth et al.\cite{ISI:A1982MW46500023} averaged over these five NO$^+$ electronic 
 states. 
 }
\end{figure}

In Fig. \ref{fig8}, photoionization cross sections and asymmetry parameters for 
the NO$^+ {w}^3 \Delta$, ${a}^3 \Sigma^+$, ${A'}^1 \Sigma^-$, ${b'}^3 \Sigma^-$ and 
 ${W}^1 \Delta$ states are shown with the averaged experimental asymmetry parameter of 
Southworth et al.\cite{ISI:A1982MW46500023} 
The cross sections for the $a$, $A'$, $b'$ and $W$ states are relatively flat above 
the thresholds and are similar to each other. The cross section for the $w$ state 
decreases slowly from the threshold to 40 eV, with a drop around 20-23 eV.  
The magnitude of cross section is the largest 
for the ${w}^3 \Delta$ state with average value of about 5 Mb. The magnitude of 
the cross sections for the ${a}^3 \Sigma^+$, ${b'}^3 \Sigma^-$ and ${W}^1 \Delta$ states is 
about 2.0-2.5 Mb, and the ${A'}^1 \Sigma^-$ state has the smallest magnitude of about 1.0 Mb. 
The asymmetry parameters of these states have remarkably similar shapes and magnitudes, 
starting from -0.8 at the thresholds and increasing monotonically to 1.2 around 40 eV.  
Our results agree well with the experimental averaged asymmetry parameter of Southworth et al. 

\subsubsection{Discussion}

The photoionization cross sections and asymmetry parameters calculated in this work 
have numerous sharp narrow peaks. They are originated from the Rydberg resonances 
associated with the excited electronic states of the molecular ions as well as two- or 
many-electron excited resonances. This can be checked by changing number of excited ionic states 
included in the R-matrix calculation or by modification of the $X_q$ terms in Eq. (\ref{eq7}). 
These sharp resonances are characteristic of multi-channel method such as present CASSCF 
multi-channel R-matrix calculation and the Schwinger multi-channel method. 
Since they are not described in the single-channel SCF target R-matrix calculation, 
the SCF cross section and asymmetry parameter in Figs \ref{fig1} and \ref{fig5} are very smooth. 
These numerous peaks are rarely seen in the previous theoretical calculations 
because many of them used single-channel method. 
The cross sections and asymmetry parameters of Stratmann 
et al.\cite{ISI:A1996UP20200016}, calculated by the Schwinger multi-channel method, 
are very similar to our CASSCF results, including location of several peaks. 
For NO photoionization, some energy regions exist where our results have 
many narrow peaks whereas the results of Stratmann et al. are smooth. 
This discrepancy is attributed to difference of NO$^+$ electronic states included 
in the calculations. For example, Stratmann et al. did not include NO$^+$ ionic states 
with $(\pi)^{-1}$ configuration, however, we put these electronic states 
in the R-matrix model. 
These narrow peaks in our results are not observed in the previous experimental 
results. In this work, we employed the fixed nuclei approximation. 
However, if we include the effect of vibrational motion by using the adiabatic 
averaging method or the non-adiabatic R-matrix method, these narrow sharp peaks may be 
averaged and look less prominent. 

As can be seen from the figures, the shapes of the photoionization asymmetry parameters 
are closely related to the main electronic configuration of the molecular ion, 
in other words, the molecular orbital from which ionization occurs. 
When the molecular ion has a $(\sigma)^{-1}$ configuration, e.g., N$_2^+ {X}^2 \Sigma_g^+$ 
and NO$^+ {b}^3 \Pi$ states, the asymmetry parameter is generally 0.5-1.0 near the threshold 
and remains nearly constant or decreases slightly with energy. In contrast, when 
the configuration is $(\pi)^{-1}$ type, e.g., N$_2^+ {A}^2 \Pi_u$ and NO$^+ {X}^1 \Sigma^+$ states, 
the asymmetry parameter is negative or zero near the threshold 
and increases rapidly with photon energy. 
These behaviour have been discussed by several authors including  
Southworth et al.\cite{ISI:A1982MW46500023} and Thiel\cite{ISI:A1983QU92400009}, 
and are interpreted in terms of angular momentum components of photoelectrons near threshold.  

In our model, angular momentum of photoelectron was considered up to $l$ = 5.  
Photoionization cross sections and asymmetry parameters obtained by maximum angular momentum 
of $l$ = 3 and 5 are very similar to each other, which suggests that we do not have to 
increase $l$ more than 5 in this study. 
Concerning number of excited electronic states of molecular ion, we included 
80 and 60 electronic states of N$_2^+$ and NO$^+$ ions, respectively. This inclusion of 
many excited ionic states is necessary to obtain converged cross sections and 
asymmetry parameters as well as to suppress pseudo resonances. The numbers of states 
in our models are much larger than those in the previous Schwinger multi-channel calculations 
of Stratmann et al.\cite{ISI:A1995RA43100054,ISI:A1996UP20200016} 
In the asymptotic region, we considered only the Coulomb potential in this work. 
In order to inspect the effect of dipole and quadrupole potentials on resonance 
position, we calculated elastic cross sections of electron - N$_2^+$ and NO$^+$ collisions 
with and without the multipole potentials, using CASSCF multi-channel R-matirx models. 
As in the photoionization cross sections shown in the figures, the elastic cross sections 
of the electron-ion collisions contain many sharp resonance peaks. 
As far as we have checked, the effect of the multipole potentials on resonance position 
in these elastic cross sections is not significant, less than 0.01 eV. 
This observation suggests that arrangement of resonances in photoionization cross section 
does not change much by introduction of the multipole potentials, at least for N$_2$ and NO 
molecules. However, when accurate assignment of Rydberg resonances and 
determination of quantum defects are of interest, 
consideration of dipole and quadrupole potentials will be important. 

In this work, we used the SA-CASSCF MOs of the molecular ion, N$_2^+$ or NO$^+$, 
to describe the ground electronic state of the neutral molecule, N$_2 {X}^2 \Sigma_g^+$ or 
NO$ {X}^2 \Pi$. 
Using a single set of MOs for both the initial and final states makes 
evaluation of transition dipole matrix elements much easier and faster than using non-orthogonal 
MO sets optimized for neutral and ionic electronic states separately. 
Our SA-CASSCF MOs are not optimized for neutral molecule, thus, the full-valence CASCI energy of the 
neutral molecule obtained by these ionic MOs is generally higher than the energy 
obtained by neutral MOs. 
Since we have extracted the energy of the neutral molecule by the R-matrix scattering calculation, 
the ground state wavefunction of the neutral molecule is constructed from the 
configuration with the valence MOs as well as the additional configurations 
which contain diffuse orbitals, as described by Burke and Taylor\cite{ISI:A1975AX03900020} 
and explained at the end of the last section. 
Inclusion of these additional configurations generally lowers the ground state energy 
compared to the full-valence CASCI energy. 
For valence electron photoionization, the effects of these ionic MOs and the additional 
configurations appear to compensate each other, since deviations of the calculated IPs 
from the experimental values are not significant as shown in Table \ref{tab1} and \ref{tab2}. 
However, in case of inner-shell ionization, this compensation does not work well, e.g.,  
the ionization threshold for the N$_2^+ (1\sigma_g)^{-1}$ state is about 4 eV lower than 
the experimental value in our preliminary calculation. Thus, we may need to use non-orthogonal MO 
sets to describe neutral and ionic states, when the R-matrix method is applied to inner-shell 
photoionization. 

Comparison of our calculation with the previous experimental and theoretical works 
demonstrates that the R-matrix method is well suited to treat valence electron 
photoionization of closed-shell as well as open-shell molecules such as N$_2$ and NO. 
We implemented the method as simple as possible. So, effects of the velocity form 
dipole moment, multipole potentials, vibrational motion etc. are not 
discussed or considered in this paper. 
In the future, we will study these issues along with application of the method 
to inner-shell molecular photoionization. 

\section{Summary}

In this work, we implemented the R-matrix method to treat molecular 
photoionization based on the procedure described by 
Burke and Taylor \cite{ISI:A1975AX03900020}. 
For the inner region calculation, the polyatomic version of the UK 
R-matrix codes was used with some modifications. 
Final state wavefunctions of molecular photoionization 
were represented by linear combinations of the R-matrix eigenstates obtained 
by diagonalization of the electronic Hamiltonian inside the inner region. 
Then, transition dipole matrix elements between the initial and the final 
states were calculated for evaluation of observables such as photoionization 
cross section and asymmetry parameter. 
As test calculations, we applied this method to valence electron photoionization 
of N$_2$ and NO molecules. 
In our R-matrix models, the electronic states of the molecular ions, N$_2^+$ and 
NO$^+$, were represented by full-valence complete active space CI wavefunctions 
with cc-pVTZ basis set. 
Calculated photoionization cross sections and asymmetry parameters 
have many narrow peaks originated from two- or many-electron excited 
resonances as well as Rydberg resonances associated with the excited 
electronic states of N$_2^+$ or NO$^+$, which are very similar to 
the previous theoretical results of Stratmann et al.\cite{ISI:A1995RA43100054,ISI:A1996UP20200016} 
obtained by the multi-channel Schwinger calculations. 
Overall shapes of the calculated cross sections and asymmetry parameters 
agree reasonably well, except for the presence of the sharp peaks, 
with the available experimental results.


\begin{acknowledgments}

The author wishes to acknowledge the helpful comments of Professors 
Keiji Morokuma and Shigeki Kato. 
\end{acknowledgments}

\clearpage


%

\clearpage

\listoffigures

\clearpage

\clearpage

%


\clearpage

\begin{table}%
\caption{\label{tab1}
  Vertical ionization potentials of N$_2$ molecule for the lower N$_2^+$ ionic states (eV).
  Experimental values are taken from Baltzer et al.\cite{ISI:A1992JX83400040}
 }
\begin{ruledtabular}
\begin{tabular}{cccc}
N$_2^+$ state & Main configuration & This work & Expt. \\
\hline
${X}^2\Sigma_g^+ $     & $(3\sigma_g)^{-1} $  & 16.06  & 15.58     \\
${A}^2\Pi_u$           & $(1\pi_u)^{-1} $     & 17.64  & 16.93     \\
${B}^2\Sigma_u^+ $     & $(2\sigma_u)^{-1} $  & 19.13  & 18.75     \\
${D}^2\Pi_g$           & $(3\sigma_g)^{-2}(1\pi_g)^{+1} $   & 25.60  & 24.79     \\
${C}^2\Sigma_u^+ $     & $(3\sigma_g)^{-1}(1\pi_u)^{-1}(1\pi_g)^{+1} $  & 26.35  & 25.51     \\
${2}^2\Pi_g$           & $(1\pi_u)^{-2}(1\pi_g)^{+1} $   & 27.06  & 26        \\
${1}^2\Sigma_u^- $     & $(3\sigma_g)^{-1}(1\pi_u)^{-1}(1\pi_g)^{+1} $   & 27.13  &           \\
${1}^2\Delta_u$        & $(3\sigma_g)^{-1}(1\pi_u)^{-1}(1\pi_g)^{+1} $  & 27.26  &           \\
${2}^2\Sigma_g^+ $     & $(2\sigma_u)^{-1}(1\pi_u)^{-1}(1\pi_g)^{+1} $  & 29.65  &           \\
\hline
\end{tabular}
\end{ruledtabular}
\end{table}

\clearpage

\begin{table}%
\caption{\label{tab2}
  Ionization potentials of NO molecule for the lower NO$^+$ ionic states (eV).
  Our IPs are calculated for the vertical ionizations, while 
  the experimental IPs taken from Albtitton et al.\cite{ISI:A1979HQ74600016} are for the 
  adiabatic ionizations. }
\begin{ruledtabular}
\begin{tabular}{cccc}
NO$^+$ state & Main configuration & This work & Expt. \\
\hline
${X}^1\Sigma^+ $     & $(2\pi)^{-1}$    &  9.29  &   9.26  \\
${a}^3\Sigma^+$      & $(1\pi)^{-1}$    & 16.13  &  15.66  \\
${b}^3\Pi $          & $(5\sigma)^{-1}$ & 16.56  &  16.56  \\
${w}^3\Delta$        & $(1\pi)^{-1}$    & 17.59  &  16.88  \\
${b'}^3\Sigma^- $    & $(1\pi)^{-1}$    & 18.23  &  17.60  \\
${A}^1\Pi $          & $(5\sigma)^{-1}$ & 18.55  &  18.33  \\
${A'}^1\Sigma^-$     & $(1\pi)^{-1}$    & 18.81  &  17.82  \\
${W}^1\Delta$        & $(1\pi)^{-1}$    & 19.10  &  18.08  \\
\hline
\end{tabular}
\end{ruledtabular}
\end{table}


\end{document}